\documentstyle[prl,aps,epsf,twocolumn]{revtex}

\title
{
Finite-Temperature Collective Excitations
 of a
  Bose-Einstein Condensate
}

\author
{
M. Olshanii
}
\address
{
Lyman Laboratory, Harvard University,
Cambridge, MA 02138, USA
{\small and}
\\
Ecole Normale Sup\'{e}rieure,
Laboratoire Kastler-Brossel,
24 Rue Lhomond,
75231 Paris Cedex 05, France  \\
{\small E-mail: {\it maxim@atomsun.harvard.edu}}
}

\date{July 30, 1998}

\begin{document}
\maketitle
\begin{abstract}
We derive a simple formula for the finite-temperature
shifts of the collective excitations of a Bose-Einstein
condensate.  
To test the validity of our treatment we apply
it to explain the
anomalous behavior of the ``$m=2$'' excitation frequency
in the recent JILA experiments (D. S. Jin {\it et al},
PRL, {\bf 78} (1997) 764), where this frequency does not approach
its ``noninteracting'' value at $\omega = 2\,\omega_{\rho}$ but, instead,
decreases with temperature.
It turns out that the effect is mainly governed by 
a resonance between the
{\it condensate} excitation frequency
and one of the quantum eigenfrequencies
of the {\it thermal cloud}.
Comparison 
of our predictions with the JILA experimental data
shows a good  agreement.
\end{abstract}
%


The recent discovery of the Bose-Einstein condensation in
dilute atomic gases \cite{BEC_first_experiments} 
has incurred a tremendous burst of  
experimental and theoretical research in this area. Whereas 
zero-temperature properties of Bose condensates seem to be 
well understood by now, theory of the finite-temperature 
behavior of the system is far from being complete. 
One of the most intriguing effects unexplained so far   
is the shifts in the spectrum of collective excitations 
at finite temperatures. A possibility of 
direct comparison of the theoretical predictions in this area 
with the experimental data \cite{Eric,Wolfgang_T_finite} 
provides us with a unique opportunity 
to test the numerous theoretical assumptions used to describe 
Bose gases in general. Among such open issues are
validity of the 
mean-field description,
relevance of the anomalous density of the non-condensed particles,
role of the many-body and trap-induced 
modifications of the interaction strength 
{\it etc}.

The first attempt to approach the problem of the finite-temperature shifts
was made in
the works \cite{Griffin_excitations,Burnett_first_excitations} 
where the collective 
excitation frequencies were associated with the 
elementary excitations spectrum of a stationary Popov Hamiltonian,
where the thermal cloud was considered as being static. In a 
subsequent paper \cite{Hutchinson_second_excitations} the model was improved 
by taking into account the many-body reduction  
of the interaction strength \cite{Stoof_T_matrix}.
The next important step forward involved the dynamical response 
of the thermal cloud \cite{Giorgini,Tosi}. 
Furthermore in the 
Ref.\cite{Stoof_excitations} the frequency shifts were calculated 
using Gaussian {\it ansatz} for both the condensate and the 
thermal cloud. Hydrodynamic treatment of the 
problem is presented in \cite{Shenoy}. Whereas all the works listed above 
were based on the mean-field treatment authors of the paper 
\cite{Shlapnikov_shifts} develop a perturbation theory 
to derive expressions for both frequencies and damping rates.

In our work we first derive a general formula for the 
finite temperature frequency shifts of the collective excitations 
of a Bose condensate: the expression we obtain is independent of
model chosen to describe the thermal cloud. 
Furthermore we notice 
that the recent finite-temperature excitation experiments 
in JILA \cite{Eric} give us an excellent opportunity to 
test our findings:  
using the fact that in the case of the ``m=2'' excitation 
a strong resonance between the 
collective excitation of interest and one of the quantum 
eigenfrequencies of the 
thermal cloud takes place, we restrict our description of the  
dynamical response of the thermal cloud to only two modes and 
obtain a simple closed set of algebraic equations, from  which we deduce
the values of the frequency 
shifts at different temperatures. We compare our 
results with the JILA experimental data.


Let us consider a time-dependent Gross-Pitaevskii-like equation
for the mean $\phi = \langle \hat{\psi} \rangle$
of the quantum field $\hat{\psi}$ of a collection of trapped  bosons 
below the Bose-Einstein transition temperature:
\begin{eqnarray}
i\hbar\frac{\partial}{\partial t} \phi =
&\left\lbrack \right.&
  \frac{p^2}{2M} + V_{\rm trap} + g(n_{c} + 2\tilde{n}) 
\left. \right\rbrack
\phi +
g\tilde{m}\, \phi^{\star} \,\,\, ;
\label{HFB_cond}
\end{eqnarray}
%
here and below 
$n_{c} = N_{c} |\phi|^2$ and $m_{c} = N_{c} \phi^2$ are the condensate
normal and ``anomalous'' densities, $N_{c}$ is the number of particles in the 
condensate,
$\tilde{n} = \langle \delta\hat{\psi}^{\dagger} \delta\hat{\psi} \rangle$ and 
$\tilde{m} = \langle \delta\hat{\psi} \delta\hat{\psi} \rangle$ are 
the normal and ``anomalous'' densities of the excited particles
expressed through the variance of the field 
$ \delta\hat{\psi} = \hat{\psi} - \langle \hat{\psi} \rangle$,
the condensate wave function is supposed to be normalized to unity
at the initial instant of time,
coupling constant $g$ is defined as $g = 4\pi\hbar^2 a / M$,
$a$ is the $s$-wave atomic scattering length, and $M$ is the atomic 
mass. Note that if the thermal densities $\tilde{n}$ and 
$\tilde{m}$ are known the above equation is exact \cite{griffin}.

We will be interested in oscillating solutions   
of (\ref{HFB_cond}) whose frequency is close to the 
frequency $\omega_{{\bf k}_{\rm mode}}$ of the  
one of the zero-temperature condensate excitations. 
We will use the following {\it ansatz} 
\begin{eqnarray}
&&\phi(t)  
\label{phi_ansatz} \\
&&\hspace{.1cm} 
          = \left\{ 
            \phi^{(0)}  
          + D\,\phi^{(-)} e^{ -i(\omega_{{\bf k}_{\rm mode}} + \epsilon)t} 
          + D^{\star}\phi^{(+)} e^{ +i(\omega_{{\bf k}_{\rm mode}} + \epsilon)t}
            \right\}  
\nonumber\\
&&\hspace{7cm} \times e^{ -i\mu \,t}
\nonumber
\\
&&\tilde{n}(t)
=  \tilde{n}^{(0)}
 + \tilde{n}^{(-)}e^{ -i(\omega_{{\bf k}_{\rm mode}} + \epsilon)t}
 + \tilde{n}^{(+)}e^{ +i(\omega_{{\bf k}_{\rm mode}} + \epsilon)t}
\label{n_ansatz}\\ 
&&\tilde{m}(t)
\label{m_ansatz}\\
&&\hspace{.3cm}
  = \left\{
           \tilde{m}^{(0)}
         + \tilde{m}^{(-)}e^{ -i(\omega_{{\bf k}_{\rm mode}} + \epsilon)t}
         + \tilde{m}^{(+)}e^{ +i(\omega_{{\bf k}_{\rm mode}} + \epsilon)t}
    \right\}
\nonumber\\
&&\hspace{7cm} \times e^{ -2i\mu \,t} 
\nonumber
\end{eqnarray}
to describe the motion of the condensate and thermal cloud. Here
$\epsilon$ is the frequency shift of interest,  
$\mu$ is the chemical potential, and $D$ is the amplitude of the condensate
oscillations. We will suppose also that the thermal densities 
$\tilde{n}$ and $\tilde{m}$ are known functions of the amplitude $D$:
\begin{eqnarray}
\tilde{n} = \tilde{n}(D) \,\,\,\, \tilde{m} = \tilde{m}(D)
\end{eqnarray}
Furthermore we will assume that
the oscillations of the thermal density are ``induced'' by the
condensate oscillations and, therefore, 
the $(+)$- and $(-)$-components of the thermal density
go to zero as the condensate oscillation amplitude $D$ goes to zero:
$
\tilde{n}^{(-,+)}(D=0) = \tilde{m}^{(-,+)}(D=0) = 0 
$.

There are two independent
small parameters present in the problem:
the ratio between 
the non-condensed and condensed densities
$\tilde{n}/n_{c}$ and the amplitude  $D$.
Let us now expand $\phi$, 
$\mu$, and $\epsilon$
into a double series with respect to both $\tilde{n}/n_{c}$ and $D$,
expand $\tilde{n}$ and $\tilde{m}$ into a simple series with respect to $D$,
and then  
insert this expansions
in the
condensate equation (\ref{HFB_cond}). 
To zeroth order in $\tilde{n}/n_{c}$ and for the infinitely 
small amplitudes of the condensate oscillations, the stationary part 
of the condensate wave function is given by the ground state 
solution $\Phi$ of the Gross-Pitaevskii equation
\begin{eqnarray}
\left\lbrack 
  \frac{p^2}{2M} + V_{\rm trap} + gN_{c}|\Phi|^2 
\right\rbrack
\Phi 
= \mu_{0} \Phi \,\,\, ,
\end{eqnarray}
where the chemical potential $\mu_{0}$ is chosen in such a way that 
the condensate wave function is properly normalized:
$\langle \Phi | \Phi \rangle = 1$. 
The oscillating part of the condensate wave function 
\begin{eqnarray}
|\phi^{(-/+)} \!\!\succ
 \doteq
\left(\begin{array}{cc}
    \phi^{(-)} \\ (\phi^{(+)})^{\star}
\end{array} \right)
\end{eqnarray}
appears only in the first order in $D$
(and zeroth order in $\tilde{n}/n_{c}$):
it is given by
\begin{eqnarray}
|\phi^{(-/+)} \!\!\succ \approx | W_{{\bf k}_{\rm mode}} \!\!\succ
\end{eqnarray}
where
$
| W_{{\bf k}_{\rm mode}} \!\!\succ
 \doteq
 (
    U_{{\bf k}_{\rm mode}} , \,  V_{{\bf k}_{\rm mode}}
 )
$
is the mode function of the collective excitation we work with. 
This mode function is represented 
by an eigenstate of the Bogoliubov-de-Gennes
Liouvillian
\begin{eqnarray}
&&{\cal L}_{0} =
\left(\begin{array}{cc}
 L_{0} & M_{0}
    \\
-M_{0}^{\star} & -L_{0}
\end{array} \right)
\label{L_0}\\
&&
L_{0} =
\frac{p^2}{2M} + V_{\rm trap} + 2gN_{c}|\Phi|^2  - \mu_{0}
\nonumber\\
&&
M_{0} =  gN_{c}\Phi^2
\nonumber
\end{eqnarray}
of an energy $E_{{\bf k}_{\rm mode}} = \hbar \omega_{{\bf k}_{\rm mode}}$.
The eigenstate $| W_{{\bf k}_{\rm mode}} \!\!\succ$
is supposed to be ``normalized''  to unity: 
$
\prec \!\!
  W_{{\bf k}_{\rm mode}} | W_{{\bf k}_{\rm mode}}
\!\!\succ
= 1 \,\, ,
$
where the bilinear form $\prec \!\! \cdot | \cdot \!\!\succ$
is defined as
\begin{eqnarray}
\prec \!\!
  W_{p} | W_{q}
\!\!\succ
\doteq
\int d^3 {\bf r} \,
\{
  U_{p}^{\star} U_{q} - V_{p}^{\star} V_{q}) \,\, .
\label{scalar_product}
\end{eqnarray}

Let us now consider the terms of 
the first order in  
$\tilde{n}/n_{c}$ and zeroth order in $D$ 
in the lhs and rhs of the condensate equation (\ref{HFB_cond}).
To this order of approximation we can calculate the 
thermal-cloud-induced correction to the chemical potential 
$\Delta\mu = \mu - \mu_{0}$, and it reads
\begin{eqnarray}
   \Delta\mu =
      \prec \!\! \Phi_{a} |
      \Delta \tilde{\cal L}^{(0)}_{\,\,D=0} |
      \sigma_3 \Phi \!\!\succ \,\, .
\label{Delta_mu}
\end{eqnarray}
Note that this correction is given by an off-diagonal
matrix element of the stationary part of the 
thermal correction to the unperturbed Liouvillian ${\cal L}^{(0)}$
\begin{eqnarray}
\Delta \tilde{\cal L}^{(0)}
=
\left(\begin{array}{cc}
 2g\tilde{n}^{(0)}          &       g\tilde{m}^{(0)}
    \\
-g(\tilde{m}^{(0)})^{\star} & -2g(\tilde{n}^{(0)})
\end{array} \right) 
\label{L^(0)} 
\end{eqnarray}
calculated between
the zero-energy ``condensate'' eigenmode 
$
|\Phi \!\!\succ
 \doteq 
   (
    \Phi , \, -\Phi^{\star}
   )
$
and a state 
$
| \Phi_{a} \!\!\succ \doteq
\sqrt{N_{c}}\frac{\partial}{\partial N_{c}}
(\sqrt{N_{c}} \sigma_{3} | \Phi \!\!\succ ) 
$
introduced in \cite{Marciek,Fixed_N_Yvan} 
(so-called ``lost eigenvector'': it is the only eigenvector of the 
operator $({\cal L}_{0})^2$ 
not present in the set of the eigenstates of ${\cal L}_{0}$ itself).  
The matrix
$
\sigma_{3} =
 \left(\begin{array}{cc}
    1 &  0
    \\
    0 & -1
\end{array} \right)
$
is the third Pauli matrix.

To the same order in 
$\tilde{n}/n_{c}$ and $D$ one can show that the correction  
to the condensate wave function $\Delta\Phi = \phi^{(0)} - \Phi$
is given by a solution of 
an inhomogeneous equation
\begin{eqnarray}
{\cal L}_{0} |\Delta\Phi \!\! \succ 
= - (\Delta \tilde{\cal L}^{(0)}_{\,\,D=0} - \Delta\mu ) \, 
                               | \sigma_3 \Phi \!\!\succ
\label{Delta_Phi}
\end{eqnarray}
and is a subject of an additional constraint
\begin{eqnarray}
\prec \!\! \Phi | \Delta\Phi \!\!\succ = 0 \,\, .
\end{eqnarray}
Here
the vector $ |\Delta\Phi \!\!\succ $ is defined as
$
|\Delta \Phi \!\!\succ
 \doteq
  (
    \Delta\Phi , \,  \Delta\Phi^{\star}
  )
$.

Consider now the terms in the equation (\ref{HFB_cond}) 
which are linear with both $\tilde{n}/n_{c}$ and $D$.
To this order 
the frequency shift we are interested in is given by
\begin{eqnarray}
&& \hbar\epsilon = \, 
           \prec \!\! W_{{\bf k}_{\rm mode}} |
             \Delta {\cal L}^{(0)}_{c} 
           + \Delta \tilde{\cal L}^{(0)}_{\,\,D=0} 
           - \Delta\mu \sigma_{3} |
             W_{{\bf k}_{\rm mode}} 
           \!\!\succ
\label{epsilon}\\
&& \hspace{3.5cm}   +
         \prec 
              \!\! W_{{\bf k}_{\rm mode}} | 
              \frac{\partial}{\partial D} 
                   \Delta \tilde{\cal L}^{(-/+)}_{\,\,D=0} |
              \sigma_3 \Phi 
         \!\!\succ
\nonumber
\end{eqnarray}
where the condensate-induced correction to the Liouvillian
(\ref{L_0}) and the oscillating part of the thermal correction to it
read
\begin{eqnarray}
&&
\Delta {\cal L}_{c} =
\label{L_c} \\
&&\hspace{.3cm}
\left(\begin{array}{cc}
   2g\lbrack \Phi^{\star}\Delta\Phi + \Delta\Phi^{\star}\Phi \rbrack
   &
   2g\Phi\Delta\Phi
   \\
   -2g\Phi^{\star}\Delta\Phi^{\star}
   &
   -2g\lbrack \Phi^{\star}\Delta\Phi + \Delta\Phi^{\star}\Phi \rbrack
\end{array} \right)  
\nonumber\\
&&
\Delta \tilde{\cal L}^{(-/+)} =
\left(\begin{array}{cc}
 2g\tilde{n}^{(-)}          &       g\tilde{m}^{(-)}
    \\
-g(\tilde{m}^{(+)})^{\star} & -2g(\tilde{n}^{(+)})^{\star}
\end{array} \right) 
\label{L^(+/-)}
\end{eqnarray}
respectively. 
The formulas (\ref{Delta_mu},\ref{Delta_Phi},\ref{epsilon}) 
defining the finite-temperature frequency shift 
of the condensate oscillation frequency is the central result of our paper.

In the derivation above
we have used a conjecture of completeness of the basis 
formed by the vectors
$
\left\{ 
  \,\,
  | \Phi \!\!\succ \,;\,
  | \Phi_{a} \!\!\succ \,;\,
  | W_{\bf k} \!\! \succ \,;\,
  | \sigma_{1} W_{\bf k}^{\star}  \!\!\succ 
  \,\,
\right\}
$
(see \cite{Fixed_N_Yvan}),
where 
the mode functions $|  W_{\bf k} \!\! \succ$ 
are 
the positive energy eigenstates of the unperturbed Liouvillian:
\begin{eqnarray}
{\cal L}_{0} | W_{\bf k} \!\! \succ = 
E_{\bf k} |  W_{\bf k} \!\! \succ 
\end{eqnarray}
and 
$
\sigma_{1} =
 \left(\begin{array}{cc}
    0 &  1
    \\
    1 &  0
\end{array} \right)
$
is the first Pauli matrix. Note that to the first order in
$\tilde{n}/n_{c}$ the thermal-cloud-induced corrections to the 
mode energies read
\begin{eqnarray}
 \Delta E_{{\bf k}}
         = \, \prec \!\!
                  W_{{\bf k}} |
                  \Delta {\cal L}^{(0)}_{c}
                  + \Delta \tilde{\cal L}^{(0)}_{\,\,D=0}
                  - \Delta\mu \sigma_{3} |
                  W_{{\bf k}}
              \!\!\succ \,\, .
\label{delta_omega_thermal}
\end{eqnarray}

To define the frequency shift (\ref{epsilon}) in an excitation experiment 
one should know the stationary part $\tilde{\cal L}^{(0)}_{\,\,D=0}$ 
of the non-condensed corrections to the condensate Liouvillian
and the ``dynamical polarizability'' 
$\frac{\partial}{\partial D} \Delta \tilde{\cal L}^{(-/+)}_{\,\,D=0}$.
In general such a calculation  would require a detailed  knowledge 
of the dynamical response of all the thermal modes  and 
it is a complicated numerical problem. However in some cases 
the calculation can be simplified: that is, for example,  the case 
of an accidental degeneracy between the condensate 
oscillation frequency $\omega_{{\bf k}_{\rm mode}}$ and one of the 
eigenfrequencies of the thermal cloud 
$\omega_{{\bf k}_{2}} - \omega_{{\bf k}_{1}}$:
\begin{eqnarray}
\omega_{{\bf k}_{\rm mode}} 
\approx \omega_{{\bf k}_{2}} - \omega_{{\bf k}_{1}} \,\, .
\end{eqnarray}
In this case the time evolution of the cloud gets dominated by 
the two resonant modes (1 and 2) and the thermal cloud can be treated in  
a kind of {\it resonant approximation}, where only two modes evolve with time. 
The latter assumption is justified for the case of the
``$m=2$'' excitation experiment at JILA where the
frequency of the mode of interest
$\omega_{{\bf k}_{\rm mode}} = \omega_{n=0, m=+2} \approx 1.45 \, \omega_{\rho}$ 
turns out to be close to the differential frequency
$
\omega_{{\bf k}_{2}} - \omega_{{\bf k}_{1}} = 
\omega^{\lbrack -\rbrack}_{n=2,m=+1} - \omega_{n=0,m=-1} \approx 1.65\, 
\omega_{\rho} 
$
(mode classification corresponds to one  
suggested in \cite{Gora_TF}).

At this point we have to chose a specific model to describe 
the evolution of the thermal cloud: in what follows we will treat 
the thermal particles using a time-dependent Hartree-Fock-Bogoliubov
approximation \cite{Giorgini}. Under this approximation the
evolution of the variance
$ \delta\hat{\psi} = \hat{\psi} - \langle \hat{\psi} \rangle$ of
the bosonic field will be governed by a mean-field equation
\begin{eqnarray}
\hbar\frac{\partial}{\partial t} \delta \hat{\psi} =
&\left\lbrack \right.&
  \frac{p^2}{2M} + V_{\rm trap} + 2g(n_{c} + \tilde{n})
\left. \right\rbrack
\delta \hat{\psi}
\nonumber \\
&+& g(m_{c} +\tilde{m}) \, \delta\hat{\psi}^{\dagger} \,\, .
\label{HFB_therm}
\end{eqnarray}
Furthermore we will suppose that prior to the excitation the system
was in a thermal equilibrium.

Following the ``resonant approximation'' program 
described above 
one can 
show that in the resonant case 
the evolution of the thermal cloud is governed by a 
simple equation for an effective 
two-level atom in a resonant field of a 
Rabi frequency 
\begin{eqnarray}
&&
\hbar\Omega_{\rm Rabi}/2 
=
   2gN_{c}
   \prec \!\!
      W_{{\bf k}_{2}} |
      \{ (\sigma_{3} \Phi) \otimes W_{{\bf k}_{\rm mode}} \} |
      W_{{\bf k}_{1}}
   \!\!\succ
   D 
\nonumber\\
&& \hspace{5.5cm} + {\cal O}(D^2)
\label{Rabi}
\end{eqnarray}
and detuning
\begin{eqnarray}
&&
\hbar\delta
= \hbar\epsilon -
  \left\lbrack
      (E_{{\bf k}_{2}} + \Delta E_{{\bf k}_{2}})
     -(E_{{\bf k}_{1}} + \Delta E_{{\bf k}_{1}})
     - E_{{\bf k}_{\rm mode}}
  \right\rbrack
\nonumber\\
&& \hspace{5.5cm} + {\cal O}(D) \,\, ,
\label{Detuning}
\end{eqnarray}
where the energy shifts $\Delta E_{{\bf k}}$ are given by
(\ref{delta_omega_thermal}) and 
the tensor product $\{ \cdot \otimes \cdot \}$ is defined as
\begin{eqnarray}
&&\{ W_{p} \otimes W_{q} \}
\\
&& \,\,\,\,
=
\left(\begin{array}{cc}
 \lbrack
 U_{p}^{\star} U_{q} +
 V_{p}^{\star} V_{q}
 \rbrack
   &
 V_{p}^{\star} U_{q}
    \\
 -U_{p}^{\star} V_{q}
   &
 -\lbrack
 U_{p}^{\star} U_{q} +
 V_{p}^{\star} V_{q}
 \rbrack
\end{array} \right)  \,\,\, .
\nonumber
\end{eqnarray}
We can  identify two distinct
classes of small collective excitations of the system:
for the excitations of the first class ($\downarrow$) the stationary part
of the thermal cloud is the same as in thermodynamic equilibrium
prior to the beginning of the excitation process; for the second class
($\uparrow$)
the populations of the two resonant modes are inverted with respect to the
thermal equilibrium.  

The thermal Liouvillian components 
(\ref{L^(0)}, \ref{L^(+/-)})   
we are interested in now read 
\begin{eqnarray}
&&\Delta \tilde{\cal L}^{(0)}_{\,\,D=0} =
\Delta \tilde{\cal L}_{\rm equilib.} 
\label{resonant_L_(0)}\\
&&\hspace{.3cm}+ 2g(N_{{\bf k}_{1}} - N_{{\bf k}_{2}}) 
\left(
  \{ W_{{\bf k}_{2}} \otimes W_{{\bf k}_{2}} \}
 -\{ W_{{\bf k}_{1}} \otimes W_{{\bf k}_{1}} \}
\right)
\nonumber\\
&&\hspace{6cm} \times 
\left( \frac{w+1}{2} \right)
\nonumber\\
&&\frac{\partial}{\partial D} \Delta \tilde{\cal L}^{(-/+)}_{\,\,D=0} =
2g(N_{{\bf k}_{1}} - N_{{\bf k}_{2}})  
\{ W_{{\bf k}_{1}} \otimes W_{{\bf k}_{2}} \}
\nonumber\\
&&\hspace{5.5cm} \times
\left( \frac{\partial}{\partial D} \rho_{21} \right) \,\, ,
\label{resonant_L_(+/-)}
\end{eqnarray}
where 
\begin{eqnarray}
w =
\left\{
 \begin{array}{c}
   -1, \,\,\, \mbox{\rm , for the ($\downarrow$)-mode}
    \\
   +1, \,\,\, \mbox{\rm , for the ($\uparrow$)-mode}
 \end{array}
\right.
\end{eqnarray}
is the population inversion;
\begin{eqnarray}
\rho_{21} =
   -\Omega_{\rm Rabi}  w  \, / \,
  2\delta
\label{rho_21}
\end{eqnarray}
is the coherence induced between mode 2 and mode 1;
$N_{\bf k} = \{\exp(E_{\bf k}/T) -1 \}^{-1}$ 
are the Bose-Einstein
occupation numbers in the thermal equilibrium;
$
\Delta \tilde{\cal L}_{\rm equilib.} =
(( 2g\tilde{n}_{\rm equilib.} ,\,  g\tilde{m}_{\rm equilib.}), \, 
 ( -g\tilde{m}_{\rm equilib.}^{\star} , \,  -2g\tilde{n}_{\rm equilib.}))
$
is the thermal cloud induced correction to the Bogoliubov-de-Gennes
Liouvillian (\ref{L_0}) calculated in the state of thermal equilibrium.
Notice that we have preserved the first order in $\tilde{n}/n_{c}$ terms in the 
denominator of the expression for the detuning (\ref{Detuning}), 
though  this seems to contradict
the expansion procedure we have chosen. In the 
{\it resonant} case, however,  the presence of this terms is well justified 
since in this case  
the mean-field shifts $\Delta E_{{\bf k}}$ and 
the frequency shift $\epsilon$ can approach 
the ``zeroth order'' splitting 
$
 E_{{\bf k}_{2}} 
-E_{{\bf k}_{1}} 
-E_{{\bf k}_{\rm mode}}
$. 

\begin{figure}
\begin{center}
\leavevmode
\epsfxsize=0.45\textwidth
\epsffile{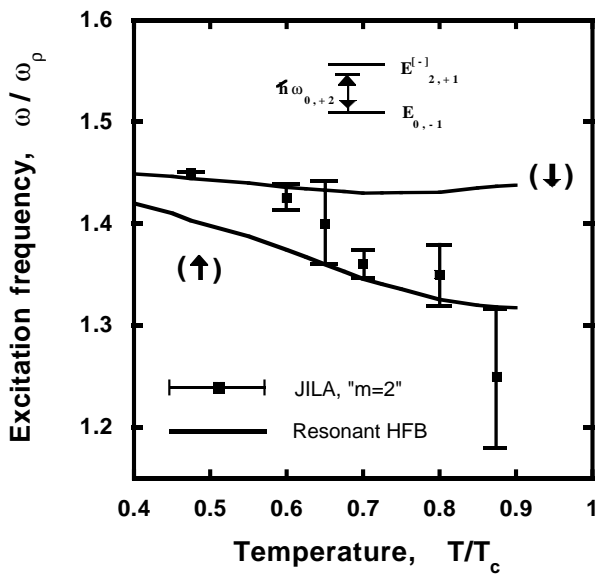}
\end{center}
\caption
{
Shift of the frequency of the ``m=2'' collective excitation 
with respect to its zero-temperature value
as
a function of temperature. Lines 
show the collective modes we have found. 
Down(up) arrow corresponds to the normal(inverted) mode. 
Errorbars show JILA
experimental data [2]. 
As a unit of frequency the 
radial trap frequency $\omega_{\rho}$ has been chosen.
The inset illustrates the relationship between the energies of the 
modes involved in the model.
\label{f_Main}
}
\end{figure}

The formula for the frequency shift (\ref{epsilon}) along 
with those for the corrections to the Bogoliubov-de-Gennes
Liouvillian (\ref{L_c},\ref{resonant_L_(0)}) and 
dynamical polarizability of the thermal cloud (\ref{resonant_L_(+/-)})
provides us with a closed implicit system of algebraic equations 
from which one can extract the values of the 
finite-temperature resonant corrections to the 
oscillations frequencies of the condensate. 
At the Fig.\ref{f_Main} we show a comparison of our predictions with 
the experimental measurement of the ``m=2'' condensate 
frequencies at JILA. We would suppose that 
in the temperature range below $T \sim .65 \, T_{c}$  
the ``normal'' mode was excited in the experiment, whereas above this 
temperature the experimental points correspond to the 
``inverted'' mode. The point at $T \sim .65 \, T_{c}$ can be associated with
any of the two modes or, probably, with a superposition of them.

In our calculations the condensate wave-function $\Phi$,
``lost vector'' $\Phi_{a}$, chemical potential $\mu$, and
thermal-cloud-induced correction to the condensate
wave-function $\Delta\Phi$ were described within the Thomas-Fermi
approximation. 
The mode functions 
$|W_{{\bf k}_{\rm mode}}  \!\!\succ = |W_{n=0, m=+2}  \!\!\succ $,
$|W_{{\bf k}_{1}} \!\!\succ = |W_{n=0,m=-1} \!\!\succ$, and 
$|W_{{\bf k}_{2}} \!\!\succ = |W_{n=2,m=+1, \lbrack -\rbrack} \!\!\succ$
were approximated by the Thomas-Fermi expressions presented in
\cite{Gora_TF}. 
For the mode energies we were using the finite-$N$ numerical
values calculated in \cite{Li_You_modes}, as the effect 
we are describing is very sensitive
to the actual value of the energy splittings.
Throughout our calculations we where neglecting
the thermal equilibrium part 
$\Delta \tilde{\cal L}_{\rm equilib.}$ of the Liouvillian as 
this type of perturbation has been shown to have no
significant effect in the frequency shifts 
\cite{Burnett_first_excitations}.

As we pointed out above, one of our collective modes corresponds to the 
inversion of populations of the thermal modes 1 and 2. 
Whether such an inversion can be reached in the experiment 
depends on details of the excitation procedure. We have checked,
however, 
that the 3\% amplitude variation of the trap frequency 
used in the JILA experiment is a strong saturating perturbation 
of the transition between the $\{n=0,m=-1\}$ and 
$\{n=2,m=+1, \lbrack -\rbrack\}$ thermal modes, and, therefore, 
the ``inverted'' mode could, indeed,  be excited during the 
excitation stage. The answer of the question on which 
particular mode will be excited  at a given temperature would 
involve a
time-dependent analysis of the whole experimental sequence:
it is a subject of our future research.

Two conclusions can be drawn from our work: 
(a) At finite temperatures 
the spectrum of the {\it collective} excitations of the 
condensate is essentially different from the spectrum of 
the {\it elementary} excitations, in contrary to the 
zero-temperature case. Comparison of the 
expressions (\ref{epsilon}) and (\ref{delta_omega_thermal}) for 
the finite-temperature corrections to the 
former and latter shows that the collective excitation shift 
contains an extra term proportional to the ``dynamical polarizability''
of the thermal cloud;
(b) Quantization of motion of the thermal quasi-particles is 
important and it can lead to significant effects, at least in the 
resonant case described in our paper. 

Note also, that in the 
off-resonant limit our results 
are  consistent with the conclusions of the work  
\cite{Giorgini}. 


I would like
to express my appreciation for many useful
discussions with M. Naraschewski, Y. Castin, M. Prentiss, 
and J. H. Thywissen.
I am grateful to the JILA group for providing me with the experimental data.
I was supported by the National Science Foundation
grant for light force dynamics \#PHY-93-12572 and by the grant {\it PAST}
of the French Government.
This work was also partially supported by
the NSF through
a grant for the Institute for Theoretical Atomic and Molecular
Physics at Harvard University and the Smithsonian Astrophysical
Observatory.
Laboratoire Kastler-Brossel is an {\it unit\'{e} de recherche de
l'Ecole Normale Sup\'{e}rieure et de l'Universit\'{e} Pierre et Marie
Curie, associ\'{e}e au CNRS}.


\end{document}